\documentclass[journal]{IEEEtran}
\usepackage{amsmath,amsthm,amsfonts,amssymb,mathrsfs}
\usepackage{cite}
\usepackage{subfigure}
\usepackage{graphicx}
\usepackage{epstopdf}
\usepackage{color}
\usepackage{url}
\usepackage{bm}
\usepackage{bbm}
\usepackage[colorlinks,linkcolor=blue,anchorcolor=blue,citecolor=blue]{hyperref}
\usepackage{braket}
\usepackage{float}
\usepackage{cuted,flushend}

\usepackage{fancyhdr}
\begin{document}
\title{Global Entanglement Distribution with Multi-mode Non-Gaussian Operations}
\author{Mingjian He, Robert Malaney, and Jonathan Green \thanks{Mingjian He and Robert Malaney are with the University of New South Wales, Sydney, NSW 2052, Australia. Jonathan Green is with Northrop Grumman Corporation, San Diego, California, USA.}}
\maketitle
\thispagestyle{fancy}
\renewcommand{\headrulewidth}{0pt}
\begin{abstract}
Non-Gaussian operations have been studied intensively in recent years due to their ability to enhance the entanglement of quantum states. However, most previous studies on such operations are carried out in a single-mode setting, even though in reality any quantum state contains multi-mode components in frequency space. Whilst there have been general frameworks developed for multi-mode photon subtraction (PS) and photon addition (PA), an important gap exists in that no framework has thus far been developed for multi-mode photon catalysis (PC). In this work we close that gap. We then apply our newly developed PC framework to the problem of continuous variable (CV) entanglement distribution via quantum-enabled satellites. Due to the high pulse rate envisioned for such systems, multi-mode effects will be to the fore in space-based CV deployments. After determining the entanglement distribution possible via multi-mode PC, we then compare our results with the entanglement distribution possible using multi-mode PS and PA.
Our results show that multi-mode PC
carried out at the transmitter is the superior non-Gaussian operation when the initial squeezing is below some threshold.
When carried out at the receiver, multi-mode PC is found to be the superior non-Gaussian operation when the mean channel attenuation is above some threshold.
Our new results should prove valuable for next-generation deployments of CV quantum-enabled satellites.

\end{abstract}

\begin{IEEEkeywords}
multi-mode state, non-Gaussian operations, entanglement distribution, channel multiplexing.
\end{IEEEkeywords}

\section{Introduction}
Non-Gaussian operations are important ingredients for various quantum information tasks, such as quantum teleportation \cite{opatrny2000improvement}, entanglement distillation \cite{browne2003driving,kurochkin2014distillation}, and quantum error correction \cite{niset2009no}. Indeed, in this latter task non-Gaussian effects are a necessary requirement due to the no-go theorem for error correction in a pure Gaussian setting.
Having the potential to boost the loss tolerance   of quantum states, non-Gaussian operations are also widely studied in  the noiseless amplifier \cite{jeffers2011optical,kim2012quantum,gagatsos2014heralded}, and quantum key distribution \cite{huang2013performance,li2016non,zhao2017improvement,lim2017continuous,guo2017performance,ma2018continuous,he2018quantum,he2019photonic,ye2019improvement}.

 At the core of non-Gaussian operations is the application of the ladder operators $\hat{a}$ and $\hat{a}^\dagger$ to quantum states, thus altering the photon number statistics. Considering single-mode states,
Photon Subtraction (PS), which applies $\hat{a}$ to a state, was first proposed in \cite{dakna1997generating} as a means of producing cat-like states.
 The operation was then utilized as an efficient way to improve the entanglement of EPR states \cite{cochrane2002teleportation,olivares2003teleportation}.
The opposite operation, Photon Addition (PA), which applies $\hat{a}^\dagger$ to a state,  has also been shown to enhance the entanglement of EPR states, at least in a lossless environment \cite{dell2007continuous,navarrete2012enhancing,zhang2013continuous}.
Studies combining PA and PS ($\hat{a}+\hat{a}^\dagger$) have also been undertaken e.g., \cite{parigi2007probing,yang2009nonclassicality,lee2011enhancing,park2012enhanced}. Such combinations further improve the entanglement but at the price of  lower efficiency.
Different from PS and PA, Photon Catalysis (PC) applies  $\hat{a}^\dagger\hat{a}$ to a state. That is, instead of adding or subtracting  photons to or from  a state, respectively, the PC operation \textit{replaces} photons from the state. It has been shown previously that PC can enhance entanglement of squeezed EPR states  \cite{hu2017continuous,guo2019continuous}. It can also partially recover the entanglement of states that suffer attenuation  \cite{ulanov2015undoing}.

For a given quantum link and initial entangled state, which  non-Gaussian operation provides the most entanglement improvement is an interesting question to ask.
It turns out that, the best non-Gaussian operation to perform is determined by various settings including, but not limited to, the level of entanglement of the entangled state, the channel conditions, and the locale (transmitter or receiver) where the non-Gaussian operation is performed.
Considering all these facts there are two interesting conclusions with regard to the PC operation in the single-mode assumption.
At the transmitter,  PC provides the most entanglement improvement when the squeezing of the EPR states is \emph{below} a certain threshold \cite{hosseinidehaj2015entanglement}.
At the receiver, PC is the best non-Gaussian operation to improve the entanglement when the channel attenuation is \emph{above} a certain threshold \cite{lee2013entanglement}.

However, the previous works discussed above are all under the assumption that each beam of the EPR state only contains a single frequency mode.
In reality, any quantum state contains multiple frequency modes - an issue of increased concern when broadband pulses of light (narrow pulses in the time domain) are utilized. In such broadband pulses,
one frequency mode of the EPR state can be entangled with many other frequency modes.
It is therefore natural to investigate whether  results derived from single mode analysis are retained for the multi-mode case, or more generally, ask what impact can non-Gaussian operations have on a multiplexed system.
To this end, a complete framework of all multi-mode non-Gaussian operations is required.
 PS and PA in the the multi-mode case have been investigated \cite{averchenko2014nonlinear,averchenko2016multimode,ra2017tomography,walschaers2017statistical,plick2018violating,walschaers2019mode}. However, currently no general framework for multi-mode PC has been established. A key aim of our study is to remedy this situation by developing the full framework for PC in the multi-mode setting. With such a new framework in place we can then fully compare the PS, PA and PC operations in the multi-mode setting. In doing this, we will be particularly interested in
 the significant potential for entanglement multiplexing gains within such a setting \cite{christ2012exponentially,usenko2014entanglement,hosseinidehaj2016cv}.

Our comparison study will be focused on Earth-satellite channels for the quantum entanglement distribution.
Recent advances  in space-based deployment of quantum communications \cite{liao2017satellite} represent a significant step forward in the creation of global-scale quantum networks. However, it is important to further study quantum communications in this context in the search for improvement in the communications set up, particularly in regard to the global distribution of entanglement - the key resource of any quantum network. Different from \cite{liao2017satellite}, in this work we will be considering Continuous Variable (CV) technology embedded in the satellite. For a review of CV quantum communications via satellite see \cite{hosseinidehaj2018satellite}.

Our contributions in this work are,
\begin{itemize}
\item
 We develop, for the first time,  a general framework for PC in the multi-mode setting.
\item
 We use this new PC multi-mode framework to investigate improvements in entanglement distribution that can be obtained via PC operations applied over Earth-satellite channels.
 \item
 We show that the PC operation can be applied to many \emph{effective} single-mode states (termed `supermodes' in what is to follow) leading to a multiplexed entanglement gain relative to pure single-mode states.
\item
Finally, we compare the performance of multi-mode PC with multi-mode PS and multi-mode PA in terms of entanglement distribution gains.
\end{itemize}

The rest of this paper is organized as follows. In Section~\ref{sec:framework} we present our framework for the multi-mode PC. In Section~\ref{sec:multistate} we apply this framework to specific multi-mode entangled states. In Section~\ref{sec:entanglement} a comparison of multi-mode PS, PA, and PC in terms of entanglement distribution over Earth-satellite channels is provided. Section~\ref{sec:conclusion} concludes our work.

\section{The General Framework of Photon Catalysis}\label{sec:framework}
\subsection{The Single-mode Case}
\begin{figure}[t]
	\centering
	\includegraphics[width=0.49\textwidth]{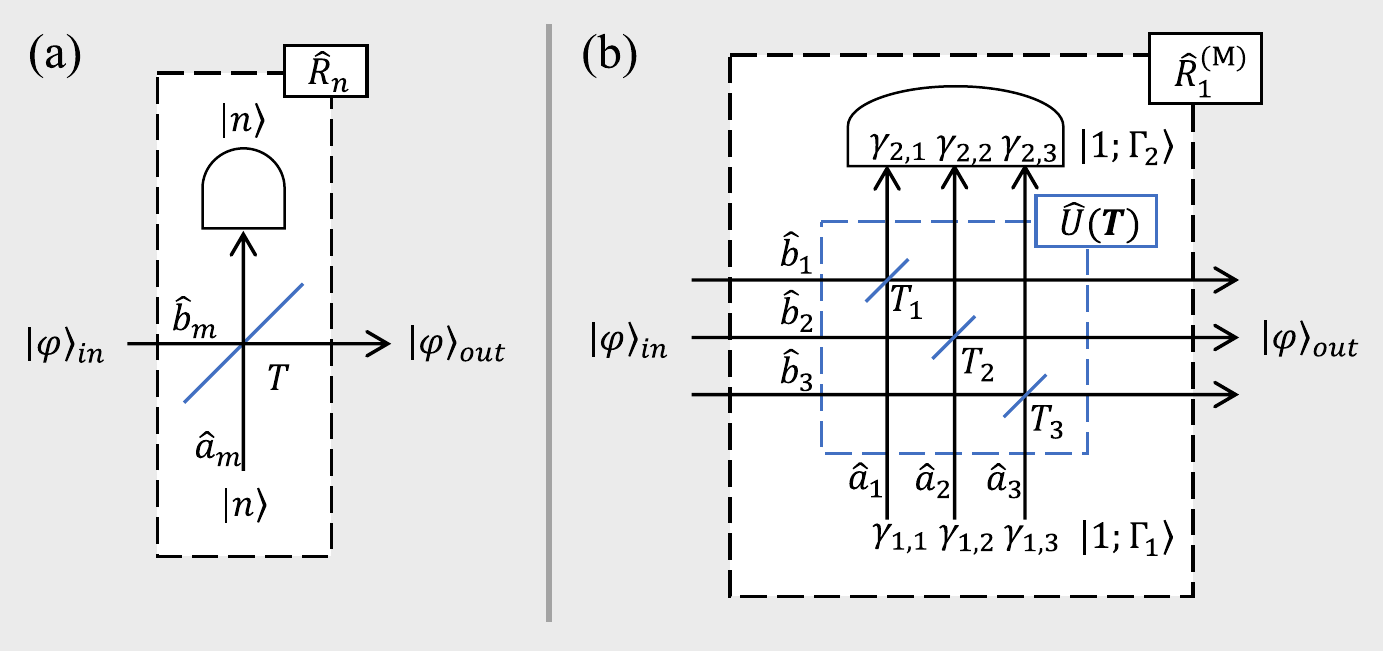}
	\caption{(a) The single-mode $n$-photon catalysis. (b) The multi-mode single-photon catalysis. Here for conciseness only three frequency modes are illustrated.}
	\label{fig:multiCatalysis}
\end{figure}

We begin with a review of the ideal single-mode photon catalysis, which will serve as the cornerstone of the multi-mode case in the next section.
Denoting $\hat{a}$ as the annihilation operator of the ancillary state of the photon catalysis operation and $\hat{b}$ the annihilation operator of the state to be catalysed (the input state), 
as illustrated in Fig.~\ref{fig:multiCatalysis}a, the single-mode $n$-photon catalysis can be described by an operator $\hat{R}_n$. In the Fock basis, this operator has the form
\begin{gather}\label{eq:singlePC}
\begin{aligned}
\hat{R}_n&=\bra{n} \hat{U}\left( T \right)\ket{n}\\
 &=\bra{0}\frac{\hat{a}^n}{\sqrt{n!}}
 \hat{U}\left( T \right) \frac{\hat{a}^{\dagger n}}{\sqrt{n!}} \ket{0},
 \end{aligned}
\end{gather}
where $\hat{U}\left( T \right)$ is the operator of a Beam Splitter (BS) with transmissivity $T$.
Using IWOP (Integration Within an Ordered Product) techniques \cite{fan2010quantum}, $\hat{U}\left( T \right)$ can be expressed as \cite{jia2014decompositions}
\begin{gather}\label{eq:BeamSplitter}
\begin{aligned}
\hat{U}\left( T \right)=&: \exp \left\{(\sqrt{T}-1)\left(\hat{a}^{\dagger} \hat{a}+\hat{b}^{\dagger} \hat{b}\right)\right.\\
&+\left.\left(\hat{a}^{\dagger} \hat{b}-\hat{a} \hat{b}^{\dagger}\right) \sqrt{\left( 1-T \right)} \right\}:,
\end{aligned}
\end{gather}
where $:\cdot:$ means simple ordering, i.e, ordering the annihilation operators to the right without taking into account the commutation relations.
A Fock state has the coherent state representation 
\begin{equation}\label{eq:UnNormedCoherent}
\ket{n}=\left.\frac{1}{ \sqrt{n !}} \frac{\partial^{n}}{\partial \alpha^{n}} \| \alpha\rangle\right|_{\alpha=0},
\end{equation}
where $|| \alpha\rangle=\exp{(\alpha \hat{a}^\dagger)\ket{0}}$. Putting Eqs.~(\ref{eq:BeamSplitter}) and (\ref{eq:UnNormedCoherent}) into Eq.~(\ref{eq:singlePC}), $\hat{R}_n$ can be expressed as \cite{hu2016multiphoton}
\begin{gather}
\begin{aligned}
\hat{R}_n=\sqrt{T}^n :L_n \left(\hat{b}^\dagger \hat{b} \frac{1-T}{T} \right):\sqrt{T}^{\hat{b}^\dagger \hat{b}},
\end{aligned}
\end{gather}
where $L_n \left(\cdot \right)$ are the Laguerre polynomials. Particularly, when $n=0$ we have the single-mode zero-photon catalysis operator
\begin{gather}\label{eq:zeroSingleCatalysis}
\hat{R}_0=\sqrt{T}^{\hat{b}^\dagger\hat{b}}.
\end{gather}
When $n=1$ we have the single-mode single-photon catalysis operator
\begin{gather}\label{eq:singCatalysis}
\hat{R}_1=\left(-\frac{1-T}{\sqrt{T}}\hat{b}^\dagger\hat{b}+\sqrt{T}\right)\sqrt{T}^{\hat{b}^\dagger\hat{b}},
\end{gather}
which can also be written as
\begin{gather}\label{eq:singCatalysisAlt}
\hat{R}_1=\left(-\frac{1-T}{\sqrt{T}}\hat{b}^\dagger\hat{b}+\sqrt{T}\right)\hat{R}_0.
\end{gather}
\subsection{The Multi-mode Case}

A multi-mode state is a superposition of single-mode states with different frequency.
Let $\hat{a}^\dagger_m$ be the creation operator of a single-mode state at a specific frequency (indexed with $m$)\footnote{In experiments $m$ will actually refer to a limited number of narrow band of frequencies, the bandwidth being set by the resolution of the detectors. Note, we do not take into account here real-world considerations of detector limitations and the ability of detectors to address individual supermodes. Instead we refer the reader to recent work in this area (e.g., \cite{brecht2011quantum,armstrong2012programmable,roslund2014wavelength}).}, the creation operator for a multi-mode state is defined as
\begin{equation}\label{eq:Agamma}
\hat{A}^\dagger_{\Gamma_k}=\sum_{m=1}^\infty \gamma_{k,m} \hat{a}^\dagger_m,
\end{equation}
where $\Gamma_k=\left[\gamma_{k,1}, \gamma_{k,2},...,\gamma_{k,\infty}\right]$ is a normalized complex vector describing the frequency profile of the multi-mode state.
The vector $\Gamma_k$ is selected from a set of vectors, $\mathbf{\Gamma} = \{\Gamma_1, \Gamma_2, ..., \Gamma_\infty\}$. 
When the vector elements of $\mathbf{\Gamma}$ are mutually orthogonal, the composed mode created by $\hat{A}^\dagger_{\Gamma_k}$ is named a \textit{supermode}. 

For the general multi-mode photon catalysis, two major components appearing in the single-mode photon catalysis need to be generalized. These are the Fock state and the BS operator.
We first define the multi-mode Fock state as
\begin{gather}\label{eq:mutliSingle}
\begin{aligned}
\ket{n;\Gamma_k}_{A}=\frac{\hat{A}^{\dagger n}_{\Gamma_k}}{\sqrt{n!}}&\ket{0},
\end{aligned}
\end{gather}
which consists of $n$ photons with the same frequency profiles.

We assume that during the multi-mode photon catalysis each single-mode component of the multi-mode state goes through separate independent BSs.
In this case, the multi-mode BS is a parallel combination of single-mode BSs. The multi-mode BS operator is defined as
\begin{equation}\label{eq:multiBS}
\hat{\textbf{U}}\left( \textbf{T} \right)=\bigotimes_{m=1}^\infty \hat{U}_m\left( T_m \right),
\end{equation}
where
$\hat{U}_m\left( T_m \right)$ is the single-mode BS operator given by Eq.~(\ref{eq:BeamSplitter}) (acting on a single-mode at the $m$-th frequency),
$\textbf{T}=\left\lbrace T_1,T_2,...,T_\infty \right\rbrace $ is the transmissivity set.

Let $\ket{\varphi}_{\text{in}}$ and $\ket{\varphi}_{\text{out}}$ be the input and output states, the multi-mode $n$-photon catalysis operation can be given by
\begin{equation}
\ket{\varphi}_{\text{out}}=\frac{\hat{R}_n^{(\text{M})}}{\sqrt{P}}\ket{\varphi}_{\text{in}},
\end{equation}
where $P$ is the success probability for the photon catalysis\footnote{In all our calculations we will assume the probability of generating the required ancillary photons is one.}, and $\hat{R}_n^{(\text{M})}$ is the multi-mode $n$-photon catalysis operator. The notation $(\cdot)^{(\text{M})}$ is used to distinguish a multi-mode non-Gaussian operator from its single-mode counterpart.
The operator $\hat{R}_n^{(\text{M})}$ has a general form
\begin{equation}\label{eq:multiphoton}
\hat{R}_n^{(\text{M})}=\left\lbrace\begin{array}{cr}
\bra{0}\frac{\prod_{k=n+1}^{2n} \hat{A}_{\Gamma_{k}}}{\sqrt{\mathcal N_2}}\hat{\textbf{U}}\left( \textbf{T} \right)\frac{\prod_{k=1}^{n}\hat{A}_{\Gamma_{k}}^\dagger}{\sqrt{\mathcal N_1}}\ket{0},&n\geq 1\\
\bra{0}\hat{\textbf{U}}\left( \textbf{T} \right)\ket{0},&n=0
\end{array}\right.
\end{equation}
where $\hat{A}_{\Gamma_{k}}^\dagger$'s are multi-mode creation operators defined in Eq.~(\ref{eq:Agamma}), and $\mathcal N_1, \mathcal N_2\in\left[ 1, n! \right]$ are normalization constants.
It is worth noting that the $n$ photons from the input ancillary Fock state might have different frequency profiles.
For the general case, the frequency profiles $\Gamma_k$ and $\Gamma_{k'}$ are not necessarily orthogonal.
The value for $\mathcal N_1$ is minimized when all the input photons have orthogonal frequency profiles. 
In this case, the input ancillary state consists of $n$ supermodes.
The value for $\mathcal N_1$ is maximized when the profiles are identical. 
In this case, the input ancillary state contains only one supermode.
Similar to $\mathcal N_1$, the value of $\mathcal N_2$ is determined by the frequency profiles of the output detected photons.

\section{Multi-mode Single-photon Catalysis (MSC)}\label{sec:multistate}
From the operational point of view, the non-deterministic property of the non-Gaussian operation is an important fact to consider in any implementation \cite{bartley2013strategies,bartley2015directly}. In this regard,
the single-photon non-Gaussian operations usually have the highest success probability for a given type of non-Gaussian operation, making them the best candidates for practical implementation. As such,
in this work we mainly focus on  single-photon non-Gaussian operations.
The multi-mode single-photon catalysis (MSC) operation is illustrated in Fig.~\ref{fig:multiCatalysis}b.
The MSC operator can be obtained by setting $n=1$ in Eq.~(\ref{eq:multiphoton}),
\begin{gather}
\begin{aligned}
\hat{R}_1^{(\text{M})}=&\bra{0}\hat{A}_{\Gamma_{2}}
\hat{\textbf{U}}\left( \textbf{T} \right)
\hat{A}_{\Gamma_{1}}^\dagger\ket{0}\\
=&\Bigg(\sum_{m,m'=1}^\infty -\gamma_{1,m} \gamma_{2,m'}^*\sqrt{1-T_m}\sqrt{\frac{1-T_{m'}}{T_{m'}}}\hat{b}_{m}^\dagger \hat{b}_{m'}\\
&+\sum_{m=1}^\infty \gamma_{1,m} \gamma_{2,m}^* \sqrt{T_m} \Bigg)\bigotimes_{m=1}^\infty \sqrt{T_m}^{\hat{b}^\dagger_m\hat{b}_m}.
\end{aligned}
\end{gather}

We are more interested in the case of a frequency non-selective beam splitter ($T_m=T$) coupled to an ancillary single-photon state that has frequency profile identical to the output photon ($\Gamma_1=\Gamma_2=\Gamma_k$), which is quite common in experiments \cite{walschaers2019mode}.
In this work, we will always use this setup when performing the non-Gaussian operations.
In this case, the MSC operator $\hat{R}_1^{(\text{M})}$ is reduced to
\begin{gather}\label{eq:general}
\hat{R}_1^{(\text{M})}=\left(-\frac{1-T}{\sqrt{T}}\hat{B}^\dagger_{\Gamma_k} \hat{B}_{\Gamma_k}  +  \sqrt{T}\right) \bigotimes_{m=1}^\infty\sqrt{T}^{ \hat{b}^\dagger_m\hat{b}_m} ,
\end{gather}
where
\begin{equation}
\hat{B}_{\Gamma_k}=\sum_{m=1}^\infty \gamma_{k,m} \hat{b}_m.
\end{equation}

For the special case when $\gamma_{1,m}=\gamma_{2,m}=\delta_{m,1}$, the MSC operator is written
\begin{gather}
\begin{aligned}
\hat{R}_1^{(\text{M})}=\left(-\frac{1-T_1}{\sqrt{T_1}}\hat{b}^\dagger_1 \hat{b}_1 + \sqrt{T_1}\right) \sqrt{T_1}^{\hat{b}_1^\dagger\hat{b}_1},
\end{aligned}
\end{gather}
which is simply the single-mode single-photon catalysis operator (on a specific single-mode) given by Eq.~(\ref{eq:singCatalysis}).


\subsection{Application of the MSC to a Specific Multi-mode Entangled State}
In the single-mode setup, an Einstein-Podolsky-Rosen (EPR) state consists of two entangled single modes. 
Likewise, in the multi-mode scenario, an EPR state consists of two entangled supermodes.
In the following , we label a supermode, $X_k$, where $X$ identifies a specific location, and $k$ identifies a specific frequency profile (from a set of orthogonal profiles).
For convenience, we introduce another two normalized vectors that characterize the frequency profiles of the two supermodes of the EPR state.
Labeling the two supermodes with $B_k$ and $D_k$, let $\Phi_k=\left[\phi_{k,1}, \phi_{k,2},...,\phi_{k,\infty}\right]$ and $\Psi_k=\left[\psi_{k,1}, \psi_{k,2},...,\psi_{k,\infty}\right]$ be the frequency profiles of $B_k$ and $D_k$, respectively, a multi-mode EPR state is defined as
\begin{equation}\label{eq:EPRsingle}
\ket{\text{EPR}_k}_{BD}=\sum_{n=0}^\infty q_{k,n} \ket{n;\Phi_k}_{B}\ket{n;\Psi_k}_{D},
\end{equation}
where 
\begin{equation}
q_{k,n} = \left(\sqrt{1-\tanh^2{r_k}}\right)\tanh^n{r_k},
\end{equation}
and with $r_k$ being the squeezing parameter,
and
\begin{gather}
\begin{aligned}
\ket{n;\Phi_k}_{B}\ket{n;\Psi_k}_{D}&=
\frac{\hat{B}^{\dagger n}_{\Phi_k}}{\sqrt{n!}}
\frac{\hat{D}^{\dagger n}_{\Psi_k}}{\sqrt{n!}} \ket{0}\\
&:=\ket{n,n;k}_{BD},
\end{aligned}
\end{gather}
where
\begin{equation}\label{eq:superAnni1}
\hat{B}^{\dagger}_{\Phi_k}=\sum_{m=1}^\infty \phi_{k,m} \hat{b}^{\dagger}_{m},
\end{equation}
and
\begin{equation}\label{eq:superAnni2}
\hat{D}^{\dagger}_{\Psi_k}=\sum_{m=1}^\infty \psi_{k,m} \hat{d}^{\dagger}_{m},
\end{equation}
are the creation operators of the two supermodes. 

Suppose we perform the MSC on $B_k$, the resultant state reads
\begin{gather}
\begin{aligned}
\ket{\psi_{\text{out}}}_{BD} =& \frac{1}{\sqrt{P}} \hat{R}_1^{(\text{M})}\ket{\text{EPR}_k}_{BD}\\
=&\frac{1}{\sqrt{P}} \left(-\frac{1-T}{\sqrt{T}}\hat{B}^\dagger_{\Gamma_k} \hat{B}_{\Gamma_k}   + \sqrt{T} \right) \\
&\times \sum_{n=0}^\infty q_{k,n} \sqrt{T}^n \ket{n,n;k}_{BD}.
\end{aligned}
\end{gather}
Suppose the ancillary single-photon state for the MSC are shaped to match the frequency profile of $B_k$, i.e., $\Gamma_k=\Phi_k$. The output state of the MSC operation can then be written
\begin{gather}
\begin{aligned}
\ket{\psi_{\text{out}}}_{BD}
&=\frac{1}{\sqrt{P}} \sum_{n=0}^\infty  q_{k,n} \sqrt{T}^{n-1} \left[ T - \left( 1-T\right)n \right]  \ket{n,n;k}_{BD}\\
&:=\frac{1}{\sqrt{P}}\ket{\text{MSC}_k}_{BD},
\end{aligned}
\end{gather}
where $\ket{\text{MSC}_k}_{BD}$ is an un-normalized multi-mode photon catalysed EPR state. 
The success probability for the MSC operation is given by
$P=\text{Tr}\{\ket{\text{MSC}_k}_{BD} \bra{\text{MSC}_k}_{BD}  \}$.
\subsection{Application of the MSC to General Multi-mode Entangled States}
More generally, the creation of EPR multi-mode entangled states arise via a Parametric Down-Conversion (PDC) process that leads to simultaneous production of multiple  EPR states, each consisting of a pair of entangled supermodes. 
In this process, a pump laser beam is first fed into a nonlinear crystal. Two correlated beams, labeled signal and idler, are then created. The resultant state can be written in the discrete form as \cite{de2014full}
\begin{gather}\label{sec:PDC}
\begin{aligned}
\ket{\mathrm{PDC}}_{BD}= \exp \left[i{\hbar}g\left( \sum_{m,m'}  L_{m,m'}\hat{b}^{\dagger}_{m} \hat{d}^{\dagger}_{m'} +\mathrm{H.c.} \right)  \right] | 0 \rangle\textrm{,}
\end{aligned}
\end{gather}
where H.c. represents the Hermitian conjugate, $g$ is the overall gain of the PDC process, $\hat{b}^\dagger_{m}$  ($\hat{d}^\dagger_{m'}$) is the creation operator of the signal  (idler) beam at the $m$-th frequency, and $L_{m,m'}$ is the Joint Spectrum Amplitude (JSA) function.
The PDC process is mainly characterized by the JSA function, which can be decomposed to the following form using the Schmidt decomposition,
\begin{equation}
L_{m,m'}=\sum_{k=1}^\infty \lambda_{k} \psi_{k,m} \phi_{k,m'}\textrm{,}
\end{equation}
where $\lambda_{k}$'s are the Schmidt coefficients, and $\psi_{k,m}$ and $\phi_{k,m'}$ are entries of the Schmidt basis $\Psi_{k}$ and $\Phi_{k}$, respectively.
The PDC state can be viewed as multiple independent multi-mode EPR states using the Schmidt decomposition. 
Specifically, the PDC state can be re-written as
\begin{gather}\label{eq:PDCstate}
\begin{aligned}
\ket{{\mathrm{PDC}}}_{BD}=&\bigotimes_{k=1}^\infty \exp \left[r_{k}  \hat{B}_{\Phi_k}^{\dagger}
\hat{D}_{\Psi_k}^{\dagger}
-\mathrm{H.c.}\right]\ket{0}\\
=&\bigotimes_{k=1}^\infty \ket{\text{EPR}_k}_{BD},
\end{aligned}
\end{gather}
where $r_k=g\lambda_k$, $\hat{B}^{\dagger}_{\Phi_k}$and $\hat{D}^{\dagger}_{\Psi_k}$ are the supermode creation operators defined in Eqs.~(\ref{eq:superAnni1})~and~(\ref{eq:superAnni2}), and $\ket{\text{EPR}_k}_{BD}$ is a multi-mode EPR state defined in Eq.~(\ref{eq:EPRsingle}).

Given a collection of multiple EPR  states (each consisting of two supermodes) we need to define a detection strategy. The detection strategy we adopt here is one in which the leading supermode (that corresponding to highest eigenvalue) has the MSC operator applied to it, while the remaining supermodes have a zero-photon catalysis operator applied to each of them.\footnote{We adopt this strategy for illustration. Other detection strategies, such as the the case where the other supermodes have no catalysis operation applied to them will lead to similar trends as those illustrated here - although details may differ. } This new operator can be written as
\begin{gather}
\begin{aligned}
\hat{\textbf{R}}^{(\text{M})}=\hat{R}_{1}^{(\text{M})}\bigotimes_{k=2}^\infty \hat{R}_{0}^{(\text{M})},
\end{aligned}
\end{gather}
where 
\begin{equation}
\hat{R}_{0}^{(\text{M})}=\bigotimes_{m=1}^\infty\sqrt{T}^{ \hat{b}^\dagger_m\hat{b}_m}
\end{equation}
is the multi-mode zero-photon catalysis operator.

Applying $\hat{\textbf{R}}^{(\text{M})}$ to the PDC state yields
\begin{gather}\label{eq:MSCMZC}
\begin{aligned}
\ket{\psi_{\text{out}}}_{BD} =& \frac{1}{\sqrt{P}} \hat{\textbf{R}}^{(\text{M})} \ket{{\mathrm{PDC}}}_{BD}\\
=&\frac{1}{\sqrt{P}} \left[ \hat{R}_{1}^{(\text{M})}\ket{\text{EPR}_1}_{BD}  \bigotimes_{k=2}^\infty \hat{R}_{0}^{(\text{M})}\ket{\text{EPR}_k}_{BD} \right]\\
=&\frac{1}{\sqrt{P}} \left[ \ket{\text{MSC}_1}_{BD}  \bigotimes_{k=2}^\infty \ket{\text{MZC}_k}_{BD}  \right],
\end{aligned}
\end{gather}
where
\begin{gather}
\begin{aligned}
\ket{\text{MZC}_k}_{BD} &= \sum_{n=0}^\infty q_{k,n} 
\underbrace{\sqrt{T}^{\sum_{m=1}^\infty \hat{b}_m \hat{b}_m^\dagger}\ket{n;\Phi_k}_{B}}_{\sqrt{T}^n \ket{n;\Phi_k}_{B} }
\ket{n;\Psi_k}_{D} \\
&= \sum_{n=0}^\infty q_{k,n} \sqrt{T}^n \ket{n,n;k}_{BD}
\end{aligned}
\end{gather}
is an un-normalized multi-mode zero-photon catalysed (MZC) state.
The success probability $P$ for creation of $\ket{\psi_{\text{out}}}_{BD}$ is written
\begin{gather}
\begin{aligned}
P=&\text{Tr}\left\{\left[ \ket{\text{MSC}_1}_{BD}  \bigotimes_{k=2}^\infty \ket{\text{MZC}_k}_{BD}  \right] \times \text{H.c.} \right\}.\\
\end{aligned}
\end{gather}

\subsection{Comparison of Non-Gaussian Operations}
For comparison purposes (in the simulations to follow), we also introduce two other multi-mode
single-photon non-Gaussian operations, namely the multi-mode
single-photon subtraction, and the multi-mode single-photon
addition. We also need to define a detection strategy for these new operations. For fairness, we assume the two comparison operations are performed to the PDC state using a detection strategy similar to that adopted for the MSC case. 
That is, we adopt a strategy in which a multi-mode single-photon
subtraction (addition) is applied to the leading
supermode, while the remaining supermodes have a zero-photon
subtraction (addition) applied to them. Considering
that the zero-photon subtraction (and the zero-photon addition)
are equivalent to the zero-photon catalysis, the operator for
the multi-mode single-photon subtraction can be written
\begin{gather}
\begin{aligned}
\hat{\textbf{S}}^{(\text{M})}=\hat{S}_{1}^{(\text{M})}\bigotimes_{k=2}^\infty \hat{R}_{0}^{(\text{M})},
\end{aligned}
\end{gather}
where
\begin{align}
\hat{S}_1^{(\text{M})}&=\bra{0}\hat{A}_{\Phi_{1}}\hat{\textbf{U}}\left( \textbf{T} \right)\ket{0}\\
&=\sqrt{\frac{1-T}{T}}\hat{B}_{\Phi_{1}} \bigotimes_{m=1}^\infty\sqrt{T}^{ \hat{b}^\dagger_m\hat{b}_m}.
\end{align}
Likewise, the operator for the multi-mode single-photon addition can be written
\begin{gather}
\begin{aligned}
\hat{\textbf{G}}^{(\text{M})}=\hat{G}_{1}^{(\text{M})}\bigotimes_{k=2}^\infty \hat{R}_{0}^{(\text{M})},
\end{aligned}
\end{gather}
where
\begin{gather}
\begin{aligned}
\hat{G}_1^{(\text{M})}&=\bra{0}\hat{\textbf{U}}\left( \textbf{T} \right)\hat{A}_{\Phi_{1}}^\dagger\ket{0}\\
&=-\sqrt{1-T}\hat{B}_{\Phi_{1}}^\dagger \bigotimes_{m=1}^\infty\sqrt{T}^{ \hat{b}^\dagger_m\hat{b}_m}.
\end{aligned}
\end{gather}


\section{Entanglement Distribution over Earth-satellite Channels}\label{sec:entanglement}
\begin{figure}[tb]
	\includegraphics[width=.3\textwidth]{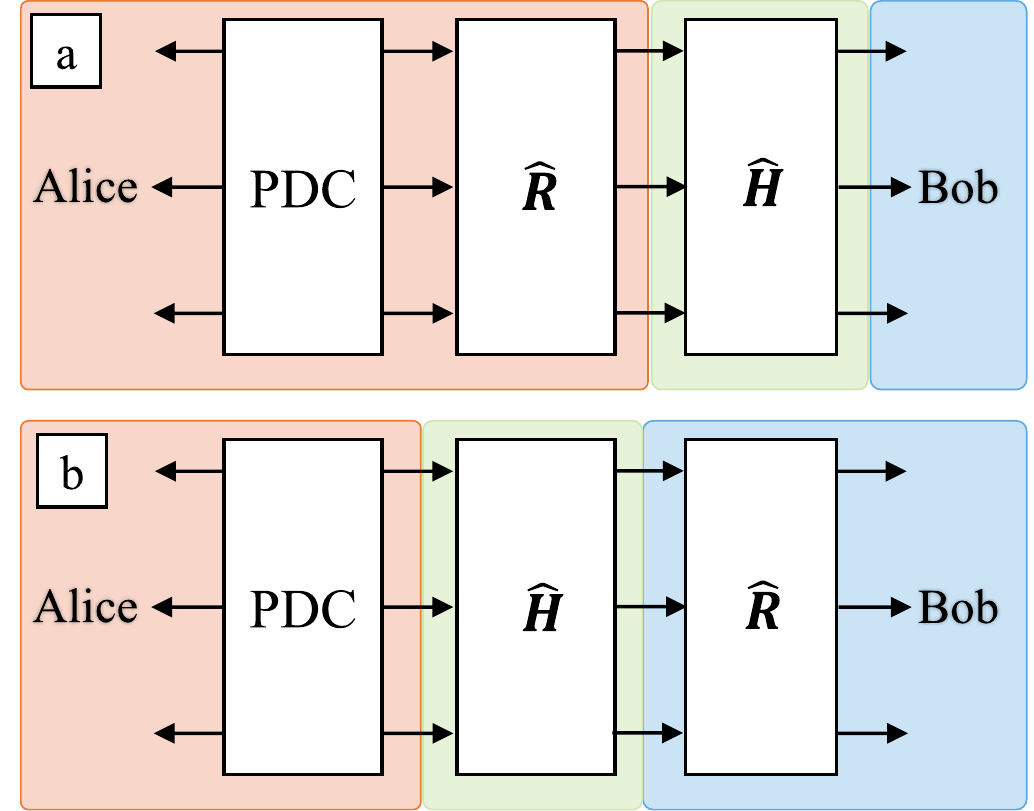}
	\centering	
	\caption{The entanglement distribution with a non-Gaussian operation at the transmitter (\textbf{a}) or at the receiver (\textbf{b}). Here the red, green, and blue boxes illustrate the transmitter, the channel, and the receiver, respectively. The channel is represented by the operator $\hat{H}$ while the non-Gaussian operation is denoted by the operator $\hat{R}$.}
	\label{fig:elnDis}
\end{figure}

In this section we study the entanglement property of the non-Gaussian operations with a concrete example, namely, the evolution of  multi-mode states over Earth-satellite channels.
In entanglement distribution, Alice (at the transmitter) will send the signal beam of the entangled state to Bob (at the receiver).
We will investigate two scenarios as shown in Fig~\ref{fig:elnDis}. In the first scenario the non-Gaussian operation is performed before the multi-mode state enters the channel.
In the second scenario, the non-Gaussian operation is performed after the state has passed through the channel.
Three types of non-Gaussian operations, namely the MSC, the multi-mode single-photon subtraction, and the multi-mode single-photon addition will be investigated.
We begin this section with a discussion of the Earth-satellite channel model.

\subsection{The Earth-satellite Fading Channels}
\begin{figure}[tb]
	\includegraphics[width=.42\textwidth]{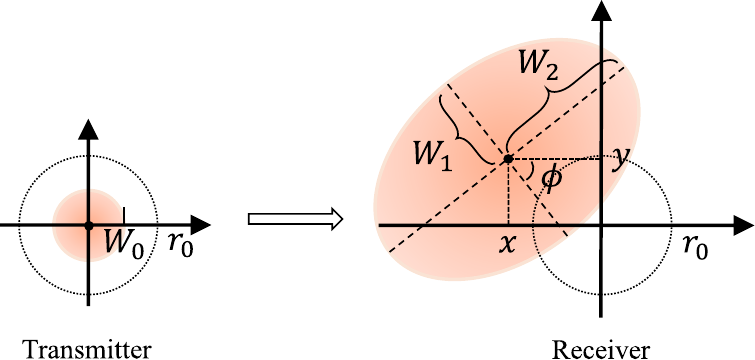}
	\centering
	\caption{Evolution of the beam-profile over the atmospheric channel. The dotted circle with radius $r_0$ illustrates the receiver aperture, and the orange eclipse represents the beam-profile at the receiver.}
	\label{fig:beam}
\end{figure}

In atmospheric channels, for optical beams the main loss mechanisms are  beam-wandering, beam-broadening, and beam-deformation, all randomly caused by turbulence in the Earth's atmosphere
\cite{andrews2005laser}.
The beam-broadening is also a consequence of diffraction.
At the receiver, the energy outside the receiver aperture is lost.
Over horizontal links the loss mechanisms are well-described by a recent model proposed by~\cite{vasylyev2016atmospheric}. In this model, the profile of the received beam is assumed to have an elliptical shape (see Fig.~\ref{fig:beam}), which is mainly characterized by five real random parameters, namely the 2-D position of the beam-centroid $(x,y)$, the semi-axes of the ellipse $W_1$ and $W_2$, and the beam rotation angle $\phi$. It is also assumed that $(x,y)$ follows a zero-mean Rayleigh distribution, $W_1$ and $W_2$ are  described by log-normal distributions,, and $\phi$ is uniformly distributed.
Let $r_0$ be the receiver aperture radius, then the channel transmissivity $\eta$ reads
\begin{equation}
\eta = \eta_0\exp\left\lbrace-\left[\frac{\sqrt{x^2+y^2}/r_0}{R(\frac{2}{W_{\rm{eff}}(\phi-\phi_0)})}\right]^{ \lambda\left(2/W_{\rm{eff}}(\phi-\phi_0)\right) }\right\rbrace\textrm{,}
\label{Cheq1}
\end{equation}
where $\phi_0=\tan^{-1}\frac{y}{x}$,
\begin{equation}
\begin{array}{*{20}{l}}	
W_{\rm{eff}}^2(\phi)= & 4r_0^2\left\lbrace\mathcal{W}\left(\frac{4r_0^2}{W_1W_2}e^{(r_0^2/W_1^2)\left[1+2\cos^2(\phi)\right]}\right.\right.\\ &
\times\left.\left. e^{(r_0^2/W_2^2)\left[1+2\sin^2(\phi)\right]}\right)\right\rbrace^{-1}
\end{array}
\end{equation}
is the squared effective spot-radius, and
\begin{equation}
\begin{array}{*{30}{l}}
\eta_0 =
& 1 - I_0\left(r_0^2\left[\frac{1}{W_1^2}-\frac{1}{W_2^2}\right]\right)e^{-r_0^2\left(1/W_1^2+1/W_2^2\right)}\\
& - 2\left[1-e^{-(r_0^2/2)\left[1/W_1-1/W_2\right]^2}\right] \\
&\times\exp\left\lbrace-\left[\frac{\frac{(W_1+W_2)^2}{\left|W_1^2-W_2^2\right|}}{R\left(\frac{1}{W_1}-\frac{1}{W_2}\right)}\right]^{\lambda\left(\frac{1}{W_1}-\frac{1}{W_2}\right)}\right\rbrace
\end{array}
\end{equation}
is the maximal attainable transmissivity achieved when $(x,y)=(0,0)$ (i.e. no beam-wandering).
In the above equations, $\mathcal{W}(\cdot)$ is the Lambert W function, $I_i(\cdot)$ is the modified Bessel function of $i$-th order, $R(\xi)$  is a scaling function given by
\begin{equation}		
R(\xi) = \left[\ln\left(2\frac{1-\exp\left[-\frac{1}{2}r_0^2\xi^2\right]}{1-\exp\left[-r_0^2\xi^2\right]I_0(r_0^2\xi^2)}\right)\right]^{-\frac{1}{\lambda(\xi)}}\textrm{,}
\end{equation}	
and $\lambda(\xi)$ is a shaping function given by
\begin{equation}
\begin{array}{*{20}{l}}
\lambda(\xi) = & 2r_0^2\xi^2\frac{\exp{[-r_0^2\xi^2]}I_1(r_0^2\xi^2)}{1-\exp\left[-r_0^2\xi^2\right]I_0(r_0^2\xi^2)} \\ &
\times
\left[\ln\left(2\frac{1-\exp\left[-\frac{1}{2}r_0^2\xi^2\right]}{1-\exp\left[-r_0^2\xi^2\right]I_0(r_0^2\xi^2)}\right)\right]^{-1}.
\end{array}
\label{Cheq5}
\end{equation}

In \cite{guo2018channel} the elliptical model is combined with a scintillation model to characterize the beam evolution over Earth-satellite channels. This requires the introduction of the
scintillation index, which can be written \cite{andrews2000scintillation}
\begin{equation}
\sigma_I^2 = \exp{\left[ \frac{0.49\sigma_R^2}{\left(1+1.11\sigma_R^{12/5}\right)^{7/6}} +  \frac{0.51\sigma_R^2}{\left(1+0.69\sigma_R^{12/5}\right)^{5/6}} \right]}-1\textrm{,}
\end{equation}
where $\sigma_R^2$ is the Rytov variance given by \cite{andrews2005laser} 
\begin{equation}
\sigma_R^2 = 2.25k^{7/6} \sec^{11/6}  {\zeta} \int_{h_0}^{h_0+L} C_n^2\left(h\right)\left( h-h_0 \right)^{5/6}\,dh,
\label{sigmaR}
\end{equation}
where $\zeta$ is the zenith angle of the satellite, $h_0$ the altitude above sea level of the ground station, and $C_n^2(h)$ the refractive index structure constant. Using the Hufnagel-Valley model, $C_n^2(h)$ is described by \cite{beland1993propagation}
\begin{align}
C_n^2(h)=&0.00594(v/27)^2(h\times10^{-5})^{10}e^{-\frac{h}{1000}}\nonumber\\
&+2.7\times 10^{-16}e^{-\frac{h}{1500}}+Ae^{-\frac{h}{100}}\textrm{,}
\label{lasteq}
\end{align}
where $v$ is the rms wind speed and $A=C_n^2(0)$.

The mean and variance of $\left\lbrace x,y,\theta_1,\theta_2\right\rbrace$ can be written
\begin{gather}
\begin{aligned}
&\left< \theta_{1/2} \right>= \ln{\left[\frac{\left(1+2.96\sigma_I^2\Omega^{5/6}\right)^2}{\Omega^2\sqrt{\left(1+2.96\sigma_I^2\Omega^{5/6}\right)^2+1.2\sigma_I^2\Omega^{5/6}}} \right]},
\\
&\left< {\Delta \theta_{1/2}}^2 \right>= \ln{\left[1+{1.2\sigma_I^2\Omega^{5/6}}/{\left(1+2.96\sigma_I^2\Omega^{5/6}\right)^{2}} \right]},
\\
&\left< \Delta \theta_1 \Delta \theta_2 \right>= \ln{\left[1-{0.8\sigma_I^2\Omega^{5/6}}/{\left(1+2.96\sigma_I^2\Omega^{5/6}\right)^{2}} \right]},
\\
&\left< \Delta x^2 \right>=\left< \Delta y^2 \right>= 0.33W_0^2\sigma_I^2\Omega^{-7/6},
\end{aligned}
\end{gather}
where $\theta_{1/2}=\ln \frac{W_{1/2}^2}{W_0^2}$, $W_0$ is the initial beam waist, $\Omega=\frac{\pi W_0^2}{L\lambda}$, $\lambda$ is the beam wavelength, and $L$ is the propagation distance of the transmitted beam.

\subsection{Non-Gaussian Operations at the Transmitter}

\begin{figure*}[tb]
	\centering
	\begin{subfigure}{
			\includegraphics[width=.40\linewidth]{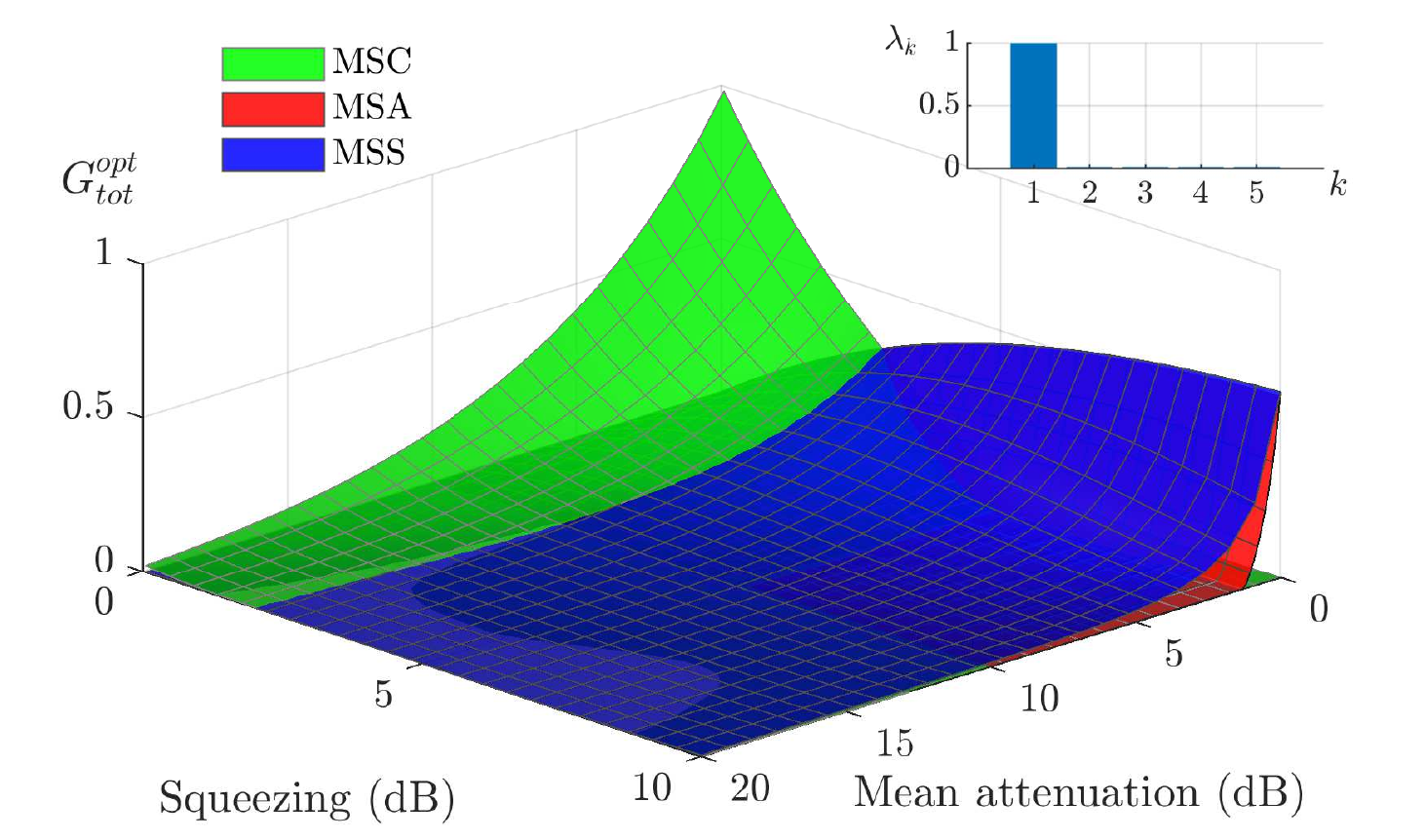}}
	\end{subfigure}
	\begin{subfigure}{
			\includegraphics[width=.40\linewidth]{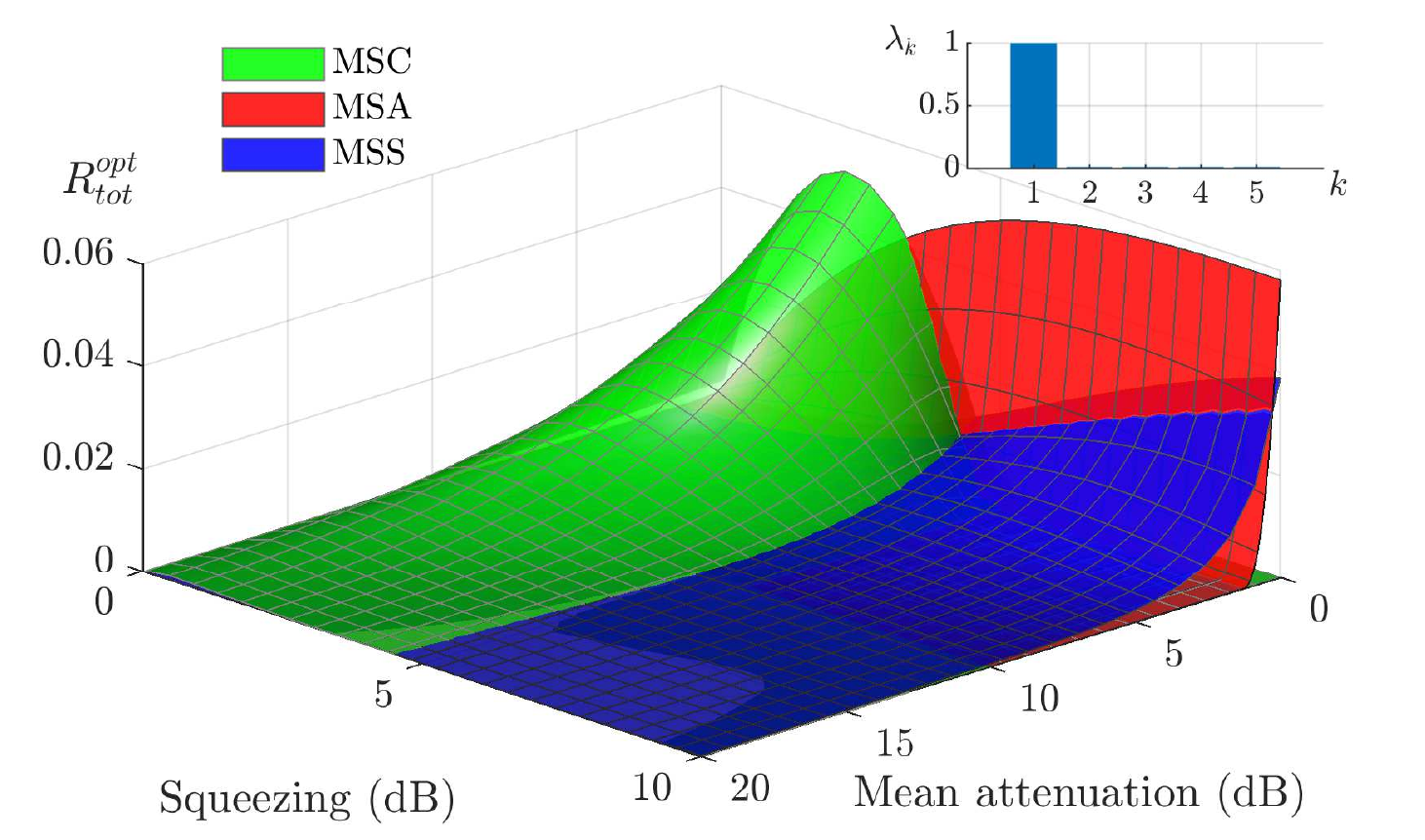}}
	\end{subfigure}	
\\
	\begin{subfigure}{
		\includegraphics[width=.40\linewidth]{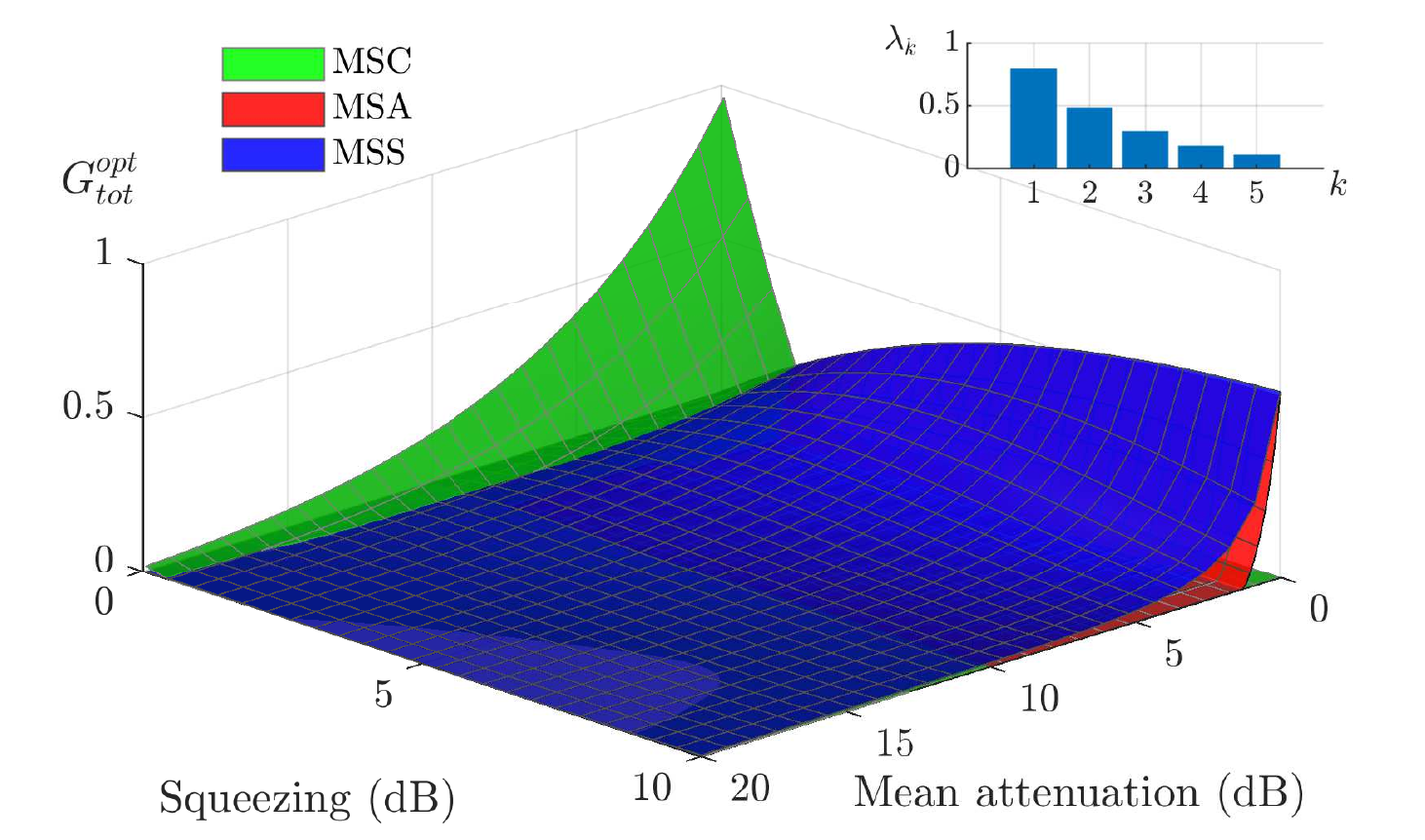}}
	\end{subfigure}
	\begin{subfigure}{
		\includegraphics[width=.40\linewidth]{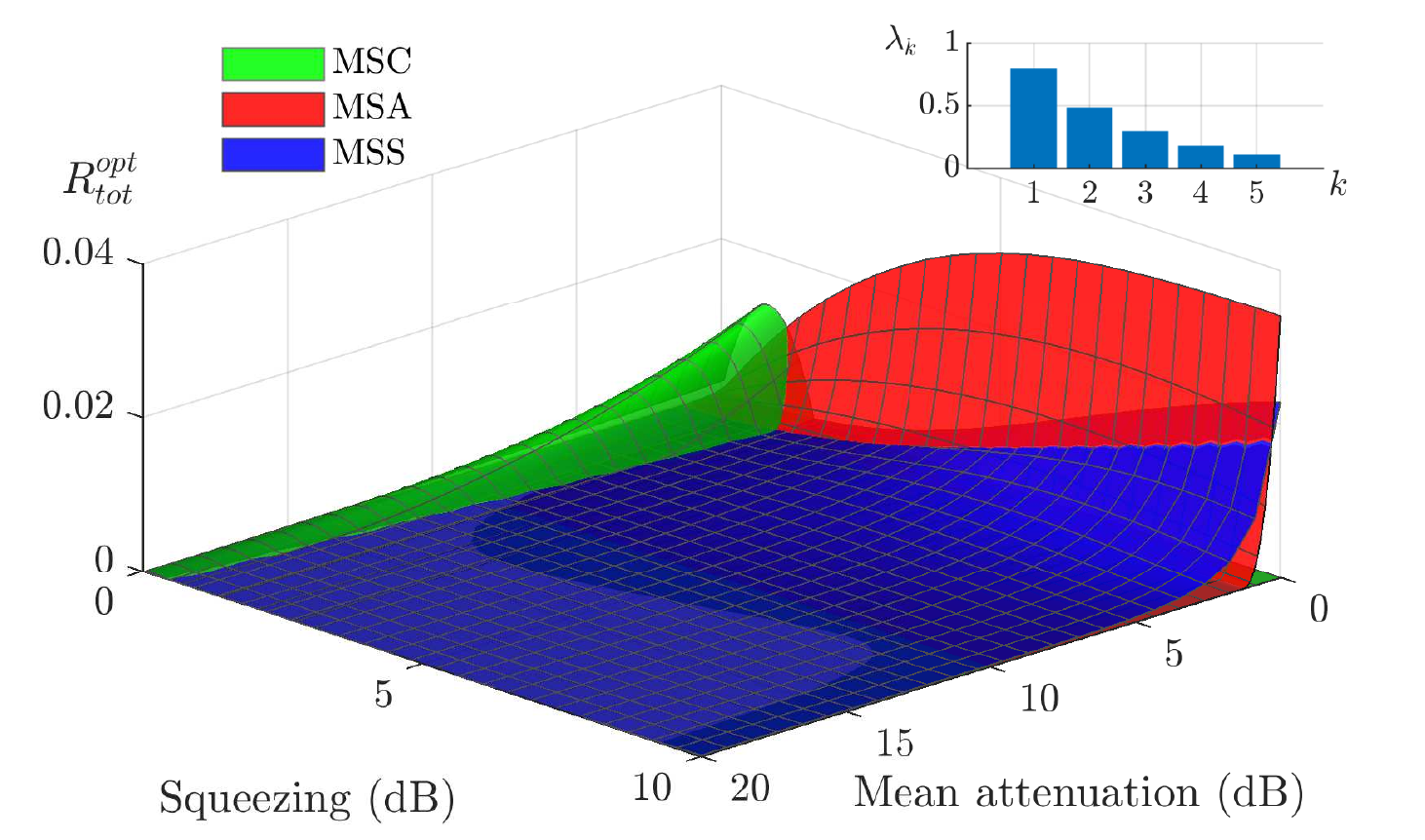}}
	\end{subfigure}
\\
	\begin{subfigure}{
		\includegraphics[width=.40\linewidth]{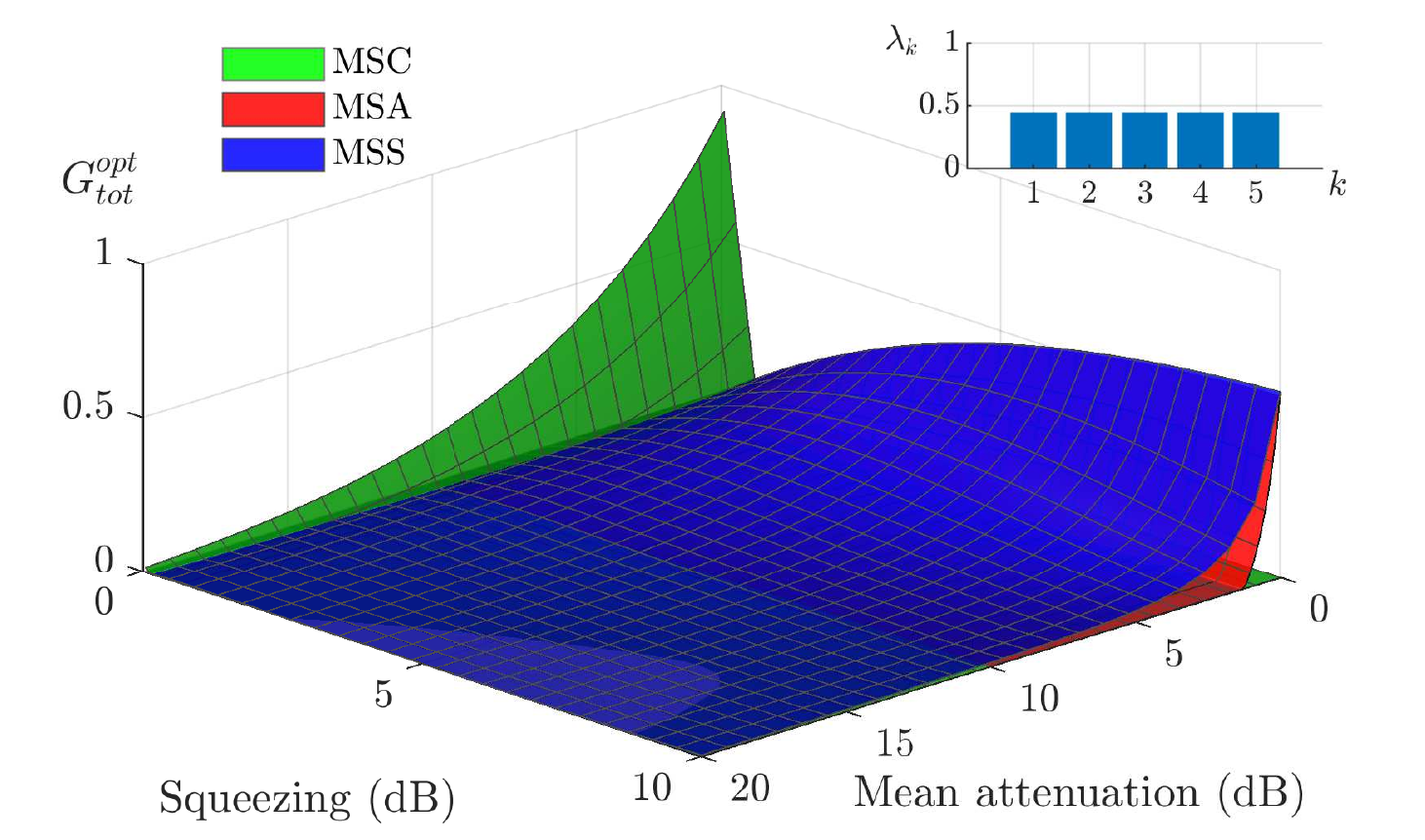}}
	\end{subfigure}
	\begin{subfigure}{
		\includegraphics[width=.40\linewidth]{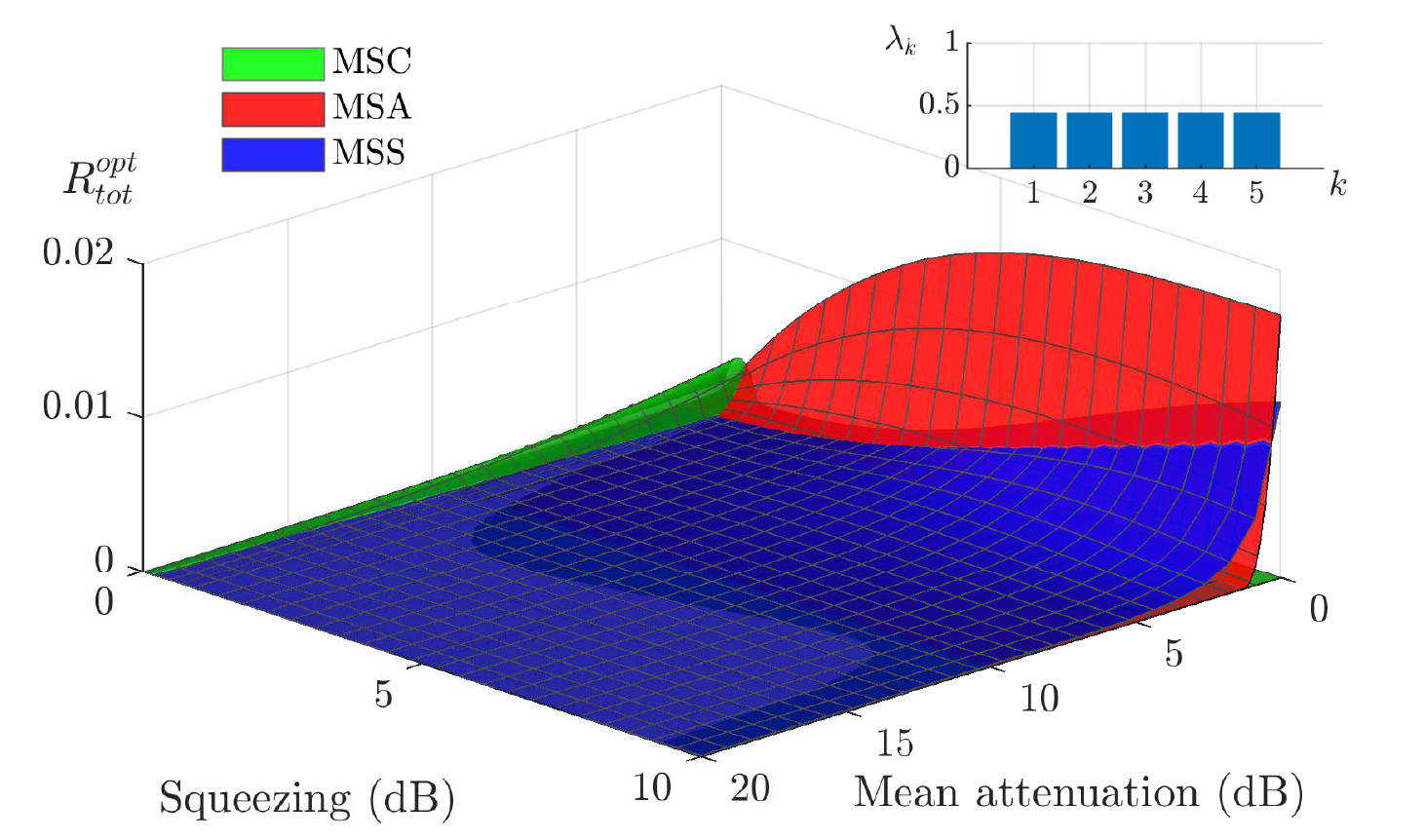}}
	\end{subfigure}
	\caption{Figures in the the left panel, from top to bottom, illustrate the optimal mean total log-negativity gain $G_{tot}^{opt}$ for the two PDC states. The insets of each figure illustrate the supermode structure of the PDC states. Figures in the right panel illustrate the optimal mean log-negativity gain rate $R_{tot}^{opt}$. Here the multi-mode non-Gaussian operations are performed at the transmitter. (MSC: multi-mode single-photon subtraction, MSA: multi-mode single-photon addition)
	}
	\label{fig:elntx}
\end{figure*}

Suppose Alice first prepares a PDC state given by Eq.~(\ref{eq:PDCstate}).
For clarity, we assume that this state only has some finite number $k_{max}$ of supermodes for each beam and Alice only performs the multi-mode non-Gaussian operation on the leading supermode of the signal beam.
Considering that the three types of non-Gaussian operations have similar derivations, we use the MSC operation as an example.
The resultant state of the MSC operation reads
\begin{gather}
\begin{aligned}
\ket{\psi}_{BD} =\frac{1}{\sqrt{P}} \left[ \ket{\text{MSC}_1}_{BD}  \bigotimes_{k=2}^{k_{max}} \ket{\text{MZC}_k}_{BD}  \right],
\end{aligned}
\end{gather}
where
\begin{gather}
\begin{aligned}
P=&\text{Tr}\left\{\left[ \ket{\text{MSC}_1}_{BD}  \bigotimes_{k=2}^{k_{max}} \ket{\text{MZC}_k}_{BD}  \right] \times \text{H.c.} \right\}.\\
\end{aligned}
\end{gather}
Alice will send the signal beam through the channel, which is modeled by a beam splitter with transmissivity $\eta_m$.
We also assume that the channel is a pure loss channel, so that the environmental mode can be modeled by a vacuum state.
Employing the IWOP techniques the channel for $B_k$ can be represented by an operator
\begin{gather}\label{eq:channelOrg}
\begin{aligned}
\hat{H}_k &= \bigotimes_{m=1}^\infty \left[\sqrt{\eta_m}^{ \hat{b}^\dagger_m\hat{b}_m} \exp\left({\sqrt{1-\eta_m}\hat{e}_m^\dagger \hat{b}_m}\right)\right] \ket{0},\\
\end{aligned}
\end{gather}
where $\hat{e}_m^\dagger$ is the creation operator of the environmental mode.
For simplicity we assume that the channel is frequency non-selective (i.e., $\eta_m=\eta$), the  channel operator can be re-written as
\begin{gather}\label{eq:channel}
\begin{aligned}
\hat{H}_k &= \left[\sqrt{\eta}^{{\hat{B}_{\Phi_k}}^\dagger \hat{B}_{\Phi_k}} \exp\left({\sqrt{1-\eta}\hat{E}_{\Phi_k}^\dagger \hat{B}_{\Phi_k}}\right)\right] \ket{0},\\
\end{aligned}
\end{gather}
where $\hat{E}_{\Phi_k}^\dagger=\sum_{m=1}^\infty \phi_{k,m} \hat{e}^{\dagger}_{m}$.
Under frequency non-selective assumption the supermodes remain orthogonal after the channel.
The entire system, $\ket{\psi}_{DBE}$, after the channel can be written
\begin{gather}\label{eq:catalysedState}
\begin{aligned}
\ket{\psi}_{BDE} =& \frac{1}{\sqrt{P}} \left[ \hat{H}_1\ket{\text{MSC}_1}_{BD}  \bigotimes_{k=2}^{k_{max}} \hat{H}_k \ket{\text{MZC}_k}_{BD}  \right]\\
=& \frac{1}{\sqrt{P}}  \sum_{n=0}^\infty \sum_{n'=0}^n q_{1,n} \sqrt{T}^{n-1} \left[ T - \left( 1-T\right)n \right] r_{n,n'}^\eta \\  
&\times \ket{n-n',n;1}_{BD}\ket{n';\Phi_1}_{E}\\ &\bigotimes_{k=2}^{k_{max}} \sum_{n=0}^\infty \sum_{n'=0}^n q_{k,n} \sqrt{T}^{n} r_{n,n'}^\eta \\
&\times \ket{n-n',n;k}_{BD}\ket{n';\Phi_k}_{E},
\end{aligned}
\end{gather}
where
\begin{gather}
\begin{aligned}
r_{n,n'}^\eta = \sqrt {\left( {\begin{array}{*{20}{c}}
		n  \\
		n'  \\
		\end{array}} \right)} {\sqrt \eta ^{n - n'}}{\sqrt {1 - \eta} ^{n'}}.
\end{aligned}
\end{gather}


\subsection{Non-Gaussian Operations at the Receiver}
\begin{figure*}[tb]
	\centering
	\begin{subfigure}{
		\includegraphics[width=.40\linewidth]{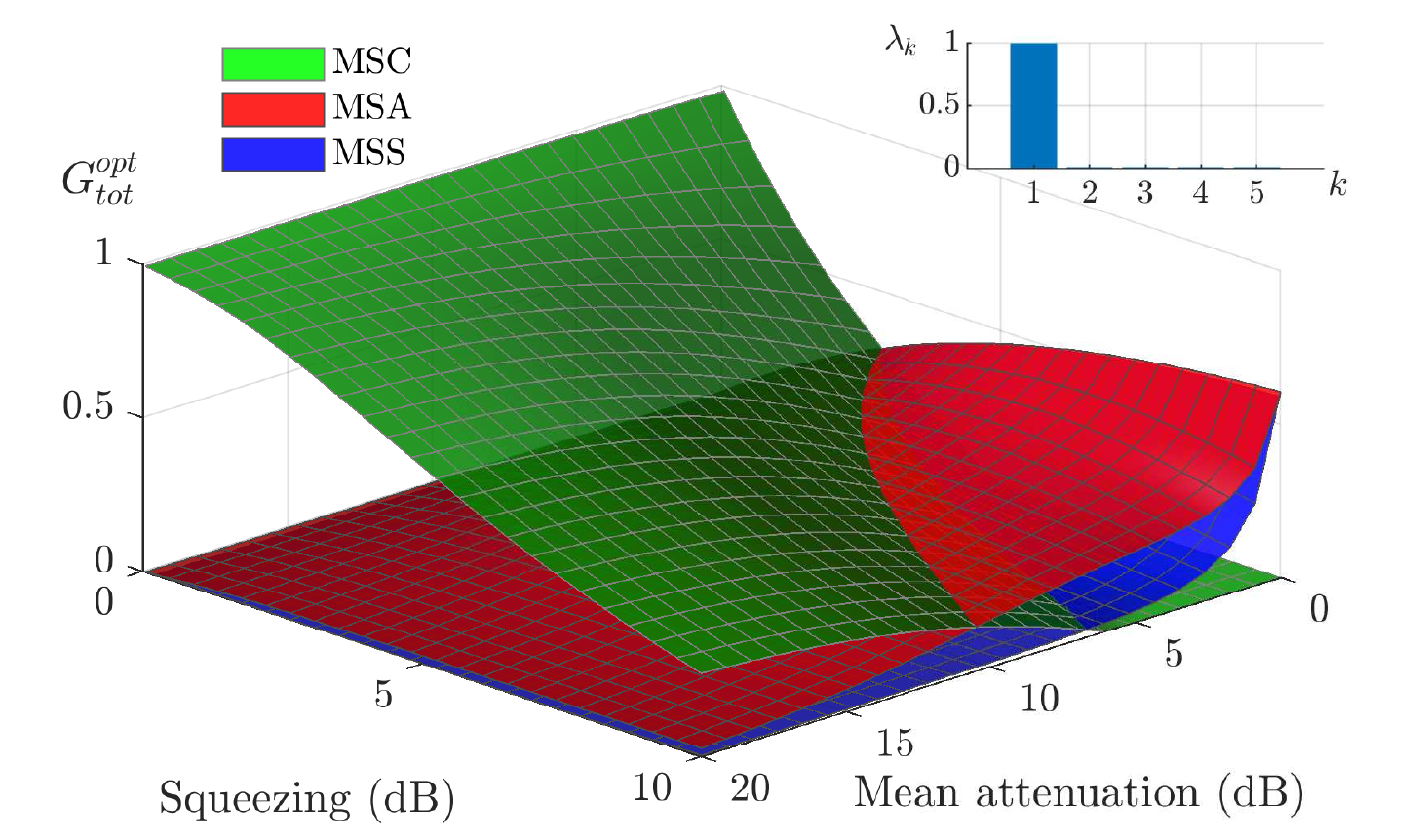}}
	\end{subfigure}
	\begin{subfigure}{
		\includegraphics[width=.40\linewidth]{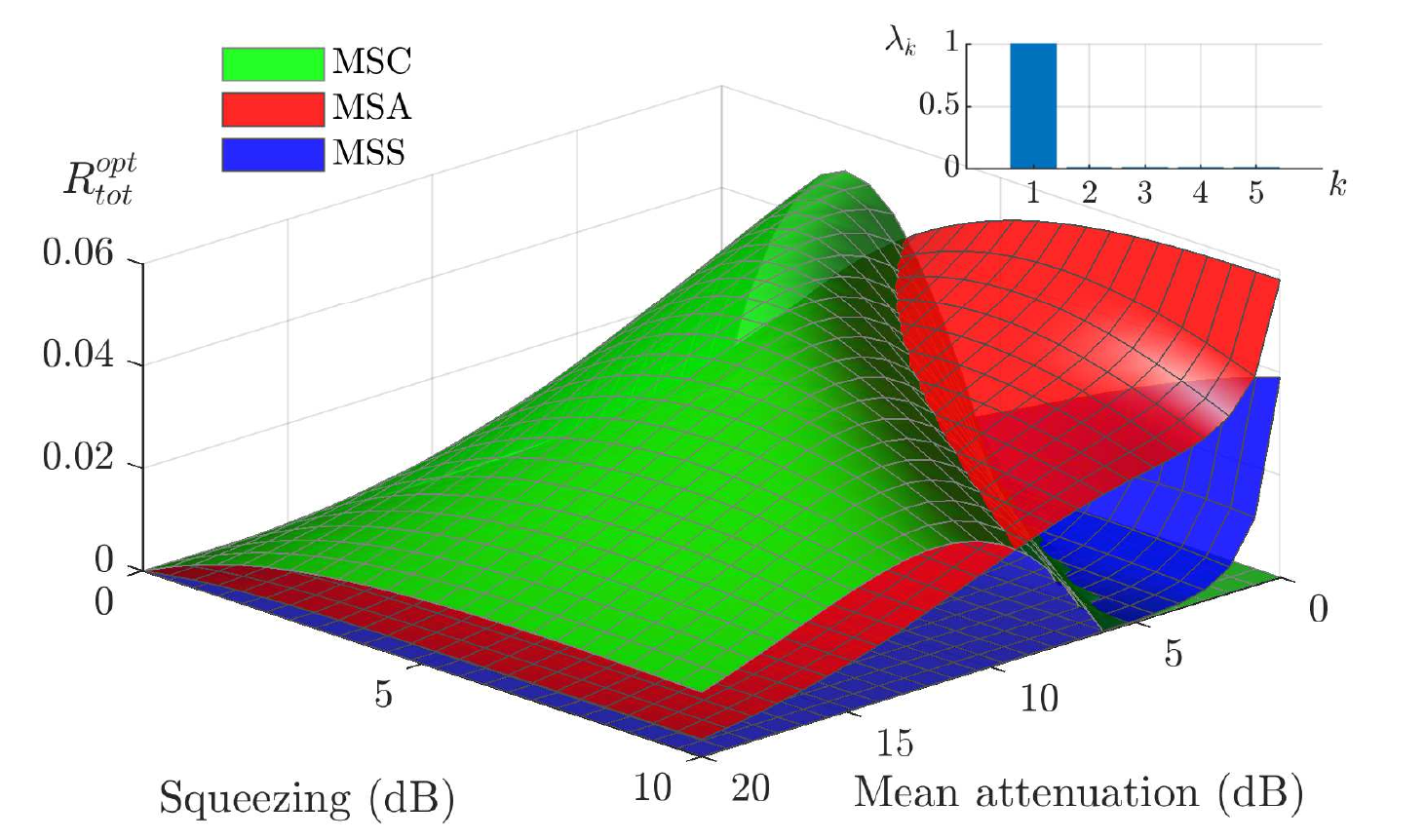}}
	\end{subfigure}
	\\
	\begin{subfigure}{
		\includegraphics[width=.40\linewidth]{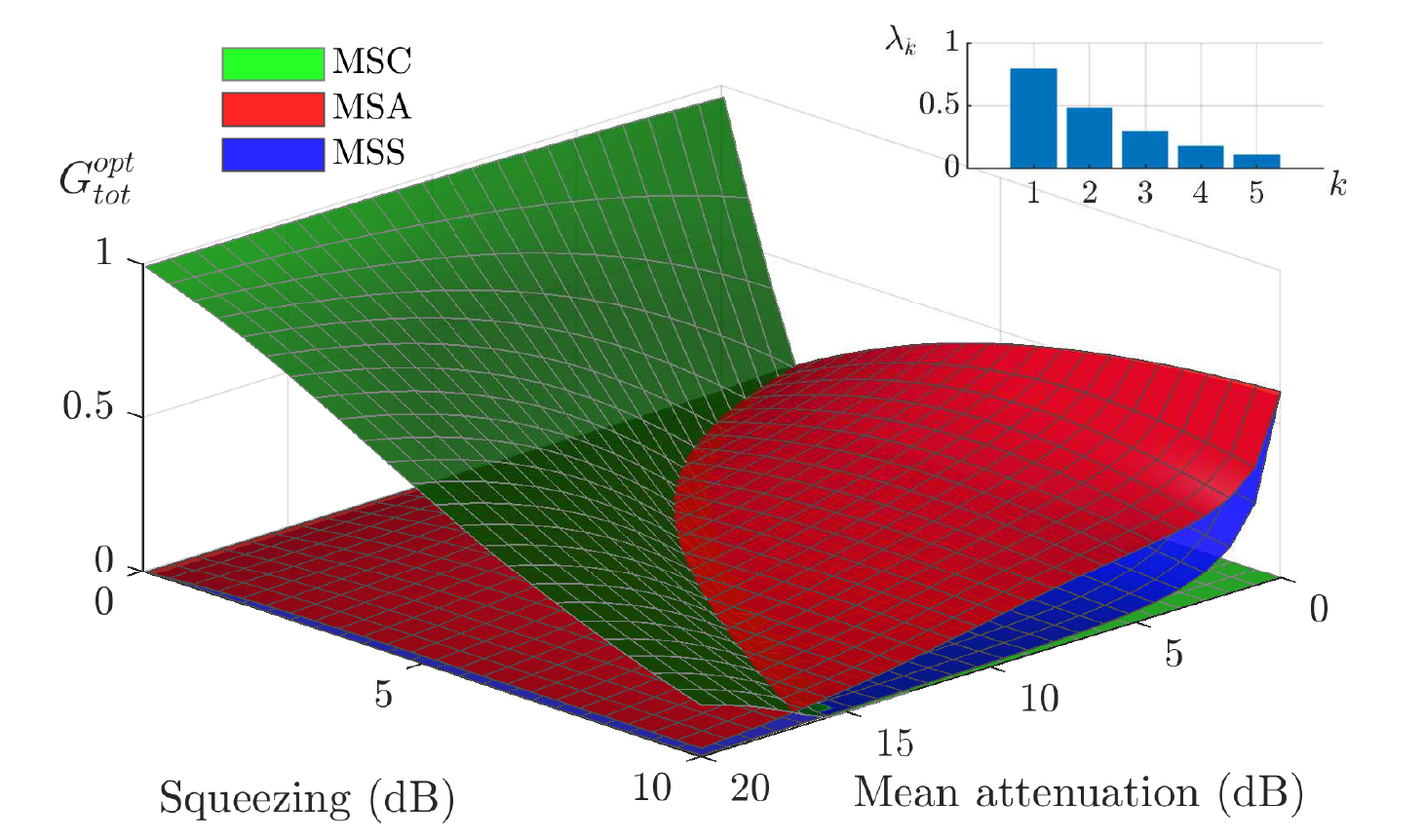}}
	\end{subfigure}
	\begin{subfigure}{
			\includegraphics[width=.40\linewidth]{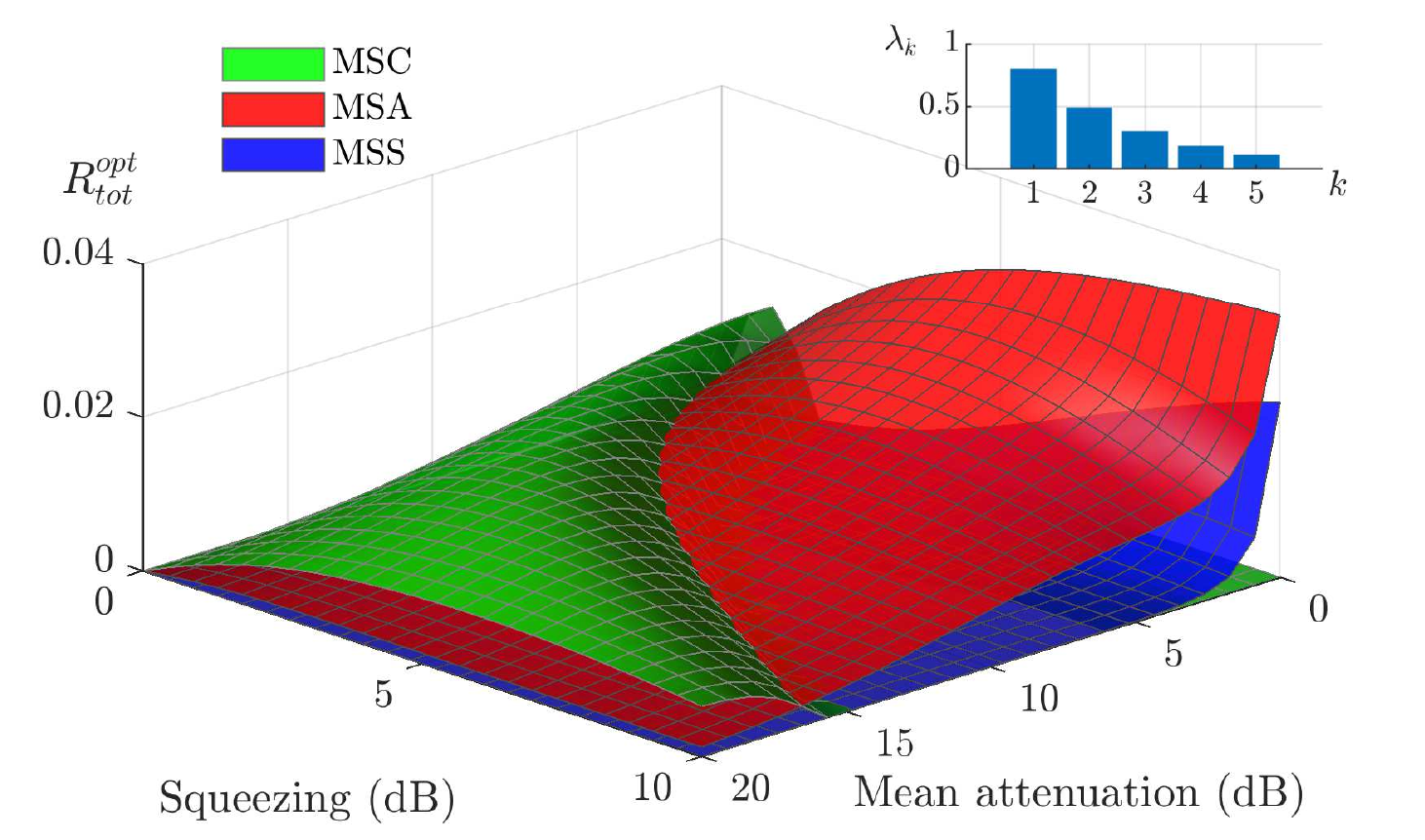}}
	\end{subfigure}	
	\\
	\begin{subfigure}{
			\includegraphics[width=.40\linewidth]{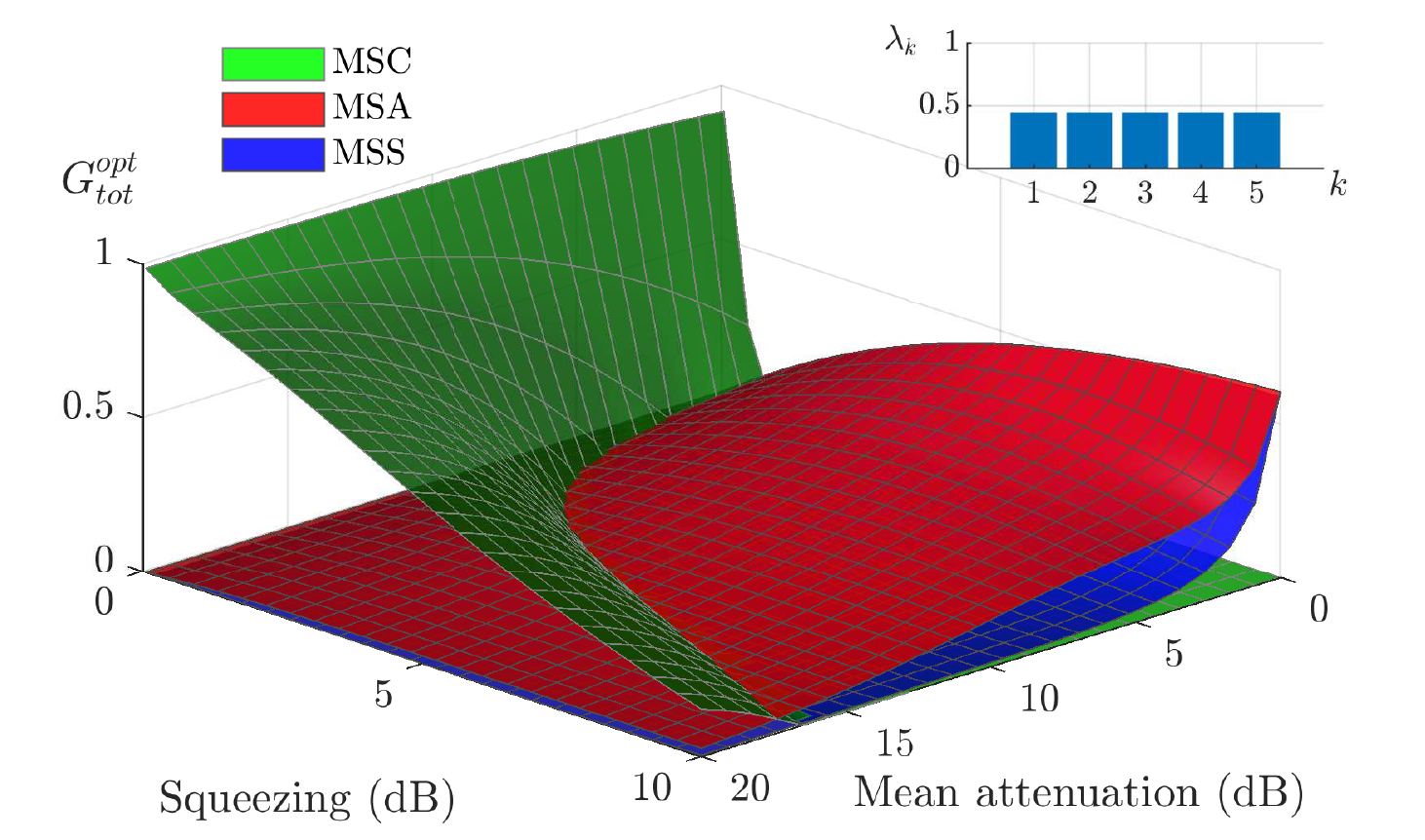}}
	\end{subfigure}
	\begin{subfigure}{
			\includegraphics[width=.40\linewidth]{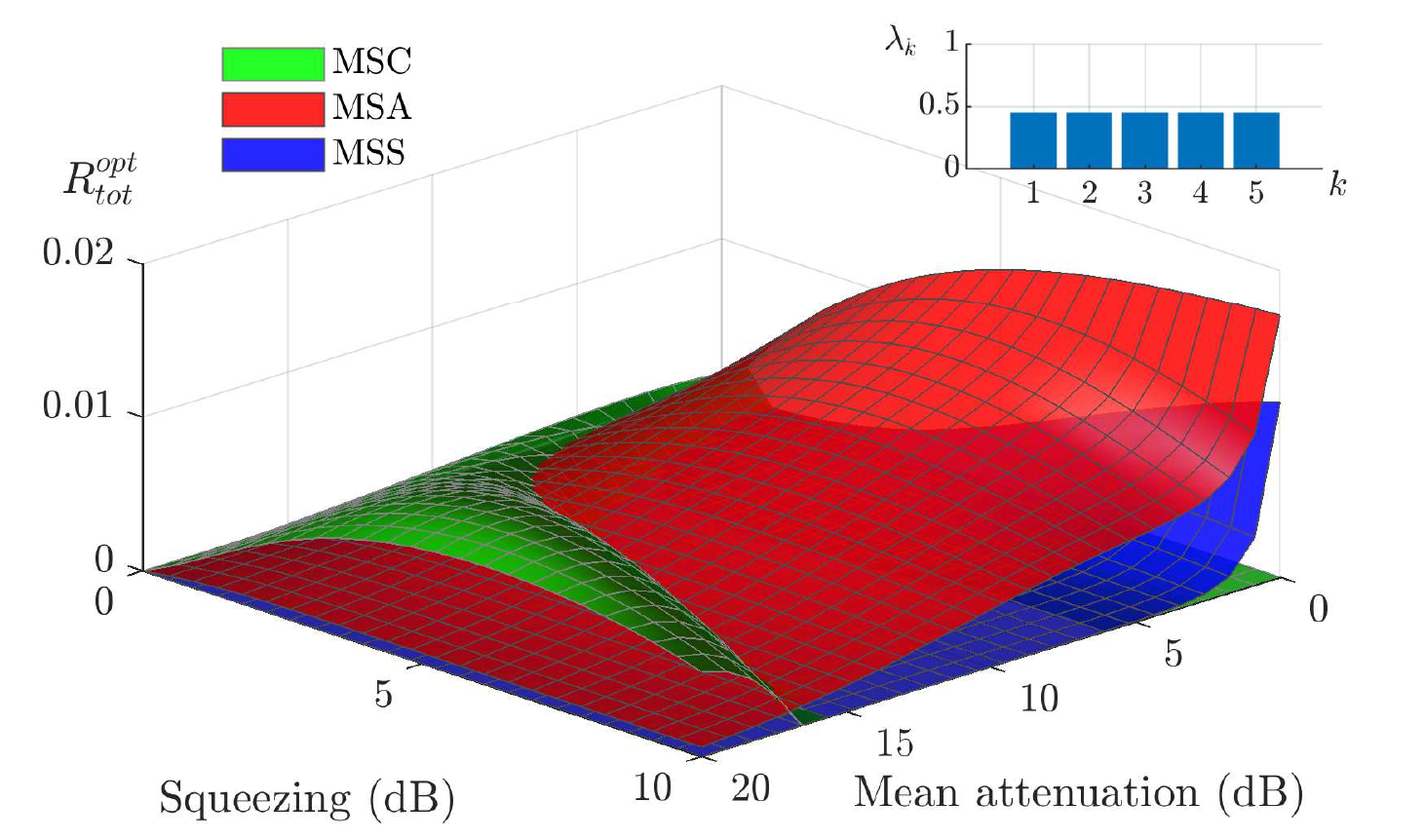}}
	\end{subfigure}	
	\caption{Same as Fig.~\ref{fig:elntx} except that the multi-mode non-Gaussian operations are performed at the receiver. (MSC: multi-mode single-photon subtraction, MSA: multi-mode single-photon addition)}
	\label{fig:elnrx}
\end{figure*}

In this section we discuss entanglement distribution with multi-mode non-Gaussian operation performed at the receiver.
In this scenario Alice first prepares a PDC state at the transmitter. She then directly sends the signal beam to Bob through the channel while keeping the idler beam. The channel alters the PDC state to the form
\begin{gather}
\begin{aligned}
&\ket{\psi}_{BDE}\\
&\quad= \bigotimes_{k=1}^{k_{max}} \hat{H}_k \ket{\text{EPR}_k}_{BD}\\
&\quad= \bigotimes_{k=1}^{k_{max}}\left[\sum_{n=0}^\infty \sum_{{n'}=0}^n q_{k,n} r_{n,{n'}}^\eta 
\ket{n-n',n;k}_{BD}\ket{n';\Phi_k}_{E}\right].
\end{aligned}
\end{gather}
After receiving the signal beam Bob will perform the MSC on this beam.
Similar to the transmitter case, we assume that Bob performs the MSC on the leading supermode of the received signal beam.
The resultant state of this operation reads
\begin{gather}
\begin{aligned}
&\ket{\psi'}_{BDE}\\
& = \frac{1}{\sqrt{P}} \sum_{n=0}^\infty \sum_{{n'}=0}^n q_{1,n} r_{n,{n'}}^\eta \sqrt{T}^{n-{n'}-1} \left[ T - \left( 1-T\right)\left(n-{n'}\right) \right] \\
&\quad\times \ket{n-n',n;1}_{BD}\ket{n';\Phi_1}_{E}\\
&\quad\bigotimes_{k=2}^{k_{max}} \sum_{n=0}^\infty \sum_{{n'}=0}^n q_{k,n} r_{n,{n'}}^\eta \sqrt{T}^{n-{n'}}\\
&\quad\times \ket{n-n',n;k}_{BD}\ket{n';\Phi_k}_{E}\\
&:= \frac{1}{\sqrt{P}} \ket{\psi''}_{BDE},
\end{aligned}
\end{gather}
where the success probability for the MSC operation reads
\begin{equation}
P=\text{Tr}\left(\ket{\psi''}_{BDE}\bra{\psi''}_{BDE}\right).
\end{equation}

\subsection{The Measurement of Entanglement}
We adopt the log-negativity $E_{LN}$ of the bipartite system as the entanglement measurement.
The log-negativity of a state $\rho$ is given by \cite{vidal2002computable}
\begin{equation}
E_{LN}(\rho)=\log_2[1+2\varepsilon(\rho)],
\end{equation}
where $\varepsilon(\rho)$ is the negativity defined as the absolute value of the sum of the negative eigenvalues of $\rho^{PT}$.
We use the total log-negativity gain, which is defined
\begin{equation}
G_{tot} = \text{max}\left[\sum_{k=1}^{k_{max}}\left( E_{LN}(\rho_k)- E_{LN}(\rho_k^G)\right),\ 0\right],
\end{equation}
as our performance metric. Here $\rho_k$ is the non-Gaussian state in the $k$-th supermode, and $\rho_k^G$ is the original Gaussian state on which the non-Gaussian operation is performed.
Since the density matrix contains infinite terms, in general the closed form solution to $E_{LN}$ is intractable.
Therefore, we approximate $E_{LN}$ by truncating the subsystems $B$, $D$, and $E$ to finite dimensions of $N_{B}$, $N_{D}$, and $N_{E}$, respectively. 
The density matrix of $B$ and $D$ is obtained by tracing out $E$ in the truncated density matrix.
This provides a lower bound to the log-negativity.

In devices which possess quantum memory the success probability, $P$, for the non-Gaussian operations can be neglected.
For the situation where the quantum memory is not applicable, we introduce another performance metric, the entanglement gain rate, which reads
$R_{tot} = P G_{tot}$.

\subsection{Simulation Results}
	For the Earth-satellite channel, we adopt the parameters from \cite{guo2018channel}. These are $h_0=0$, $v=6\text{m/s}$, and $C_n^2(0)=9.6\times10^{-14}\text{m}^{-2/3}$.
We focus on vertical down-link channels, which can be easily generalized to non-vertical cases.
The initial beam waist and the receiver aperture are set to $W_0=6\text{cm}$ and $r_0=1\text{m}$, respectively.
We note that such a receiver aperture is realistic for ground stations.
In accordance with the experiment in \cite{roslund2014wavelength}, the center wavelength of the beam is set as $\lambda=795\text{nm}$ with a bandwidth of roughly $20\text{nm}$.
We assume that all the frequency components undergo the same attenuation as the central frequency component.
In this case, both the supermode structure and orthogonality of the PDC state after the channel are retained.

A Monte Carlo algorithm is used to simulate the channel since no closed-form solutions to the PDF of the channel transmissivity $\eta$ exists. Let $\eta_s$ be the samples generated by the Monte Carlo algorithm, during state transmission, we assume that the channel transmissivity can be measured for each coherent time window, so that the transmissivity $T$ for the beam splitter for the non-Gaussian operations can be adjusted to optimize $G_{tot}$ or $R_{tot}$ for every given $\eta_s$.
We label the mean optimized $G_{tot}$ and $R_{tot}$ by $G_{tot}^{opt}$ and $R_{tot}^{opt}$, respectively.

Assuming that $k_{max}=5$, we consider three PDC states that have different supermode structures.
For the first PDC state, the energy of the supermodes, other than the leading supermode, is assumed close to zero. This state is a good approximation to a single-mode state.
The second PDC state is more realistic and contains more than one non-trivial supermodes.
For the last PDC state, we assume the 5 supermodes have the same level of squeezing.

We first investigate the scenario where the non-Gaussian operations are performed at the transmitter.
In the left panel of Fig.~\ref{fig:elntx}, $G_{tot}^{opt}$ is shown as a function of the mean channel attenuation
$\bar{\eta}\left[dB\right]=-10 \log_{10} \bar{\eta}$ and the squeezing level of the leading supermode
$r_1\left[dB\right]=-10\log_{10} \left[\exp \left(2{g\lambda_1}\right)\right]$, where $g$ is the PDC gain in Eq.~(\ref{sec:PDC}).
We find that the first conclusion mentioned in the introduction is retained for the multi-mode case. That is, at the transmitter the MSC provides larger $G_{tot}^{opt}$ when the initial squeezing is below a threshold.
In the multi-mode case this threshold is different from the single-mode case, and is determined by the supermode structure of the state. As such, the parameter settings under which a specific non-Gaussian operation maximizes entanglement is quite different in the single-mode and multi-mode cases.
In the right panel of Fig.~\ref{fig:elntx}, we compare $R_{tot}^{opt}$ for the three non-Gaussian operations.
For the cases where quantum memory is not applicable, MSC is still the best non-Gaussian operation to increase the entanglement when the initial squeezing is below a certain threshold.

We then study the scenario where the non-Gaussian operations are performed at the receiver.
Similar to Fig.~\ref{fig:elntx}, in the left panel of Fig.~\ref{fig:elnrx}, $G_{tot}^{opt}$ is illustrated as a function of $\bar{\eta}\left[dB\right]$ and $r_1\left[dB\right]$.
We find that the second conclusion mentioned in the introduction is also retained. That is, at the receiver, the MSC provides larger $G_{tot}^{opt}$ when $\bar{\eta}$ is above a certain threshold (similar for $R_{tot}^{opt}$ as illustrated in the right panel of Fig.~\ref{fig:elnrx}).
Note that the threshold value is a function of the supermode structure of the state.


\section{Conclusions}\label{sec:conclusion}
In this work we have established, for the first time, a framework for multi-mode photon catalysis operations. Using our new framework, and focusing on higher success-rate single-photon operations, we then compared photon catalysis operation with photon subtraction and photon addition operations when applied to Earth-satellite channels. Our results show that the outcomes are dependent on whether the operations are carried out at the transmitter or the receiver. In the former case we find that multi-mode photon catalysis is the superior non-Gaussian operation when the initial squeezing is below some threshold. In the latter case we find that multi-mode photon catalysis is the superior non-Gaussian operation when the mean channel attenuation is above some threshold.
Our new results will be important for next-generation deployments of quantum-enabled satellites that deploy CV technology. This will be particularly the case when on-board quantum memory becomes available.

\raggedbottom


\end{document}